\documentclass{article}
\usepackage{arxiv}
\usepackage[utf8]{inputenc} 
\usepackage[T1]{fontenc}    
\usepackage{hyperref}       
\usepackage{url}            
\usepackage{booktabs}       
\usepackage{amsfonts}       
\usepackage{nicefrac}       
\usepackage{microtype}      
\usepackage{graphicx}
\usepackage[numbers]{natbib}
\usepackage{doi}
\usepackage{xcolor}
\usepackage{float}
\newcommand{\aod}{AoD}

\title{Anti OS Fingerprinting, Revisited: A Loadable Kernel Module for FreeBSD and What Twenty-Five Years Changed}

\author{ \href{https://orcid.org/0000-0003-0967-9040}{Daisuke Miyamoto} \\
\thanks{d-miyamoto@grips.ac.jp}
	National Graduate Institute \\
    for Policy Studies\\
	Tokyo, Japan  \\
}

\renewcommand{\shorttitle}{Anti OS Fingerprinting, Revisited}

\hypersetup{
pdftitle={Anti OS Fingerprinting, Revisited: A Loadable Kernel Module for FreeBSD and What Twenty-Five Years Changed},
pdfauthor={Daisuke Miyamoto},
pdfkeywords={OS fingerprinting, Deception, Cyber resilience, FreeBSD, Kernel module},
}

\begin{document}
\maketitle

\begin{abstract}
In 2001 our research group published a paper on an Anti OS Detection (AoD)
system: a modified FreeBSD~4.3 kernel that makes a host answer OS
fingerprinting probes as if it ran another OS. This report revisits that
work on FreeBSD~15.1. I claim no new method. I report three things. First,
the same goal is reached by a loadable module of about 1,700 lines of C that
uses the packet filter hook interface and sets one kernel variable, without
changing the kernel source; it works over IPv4 and IPv6, and one setting
selects the personality. Second, a personality built after a Red Hat
Enterprise Linux~9.9 host and one built after a Windows~11 host received the
same verdicts as those hosts from Nmap~7.98 over IPv4 and IPv6, with the
same p0f~3 signatures and the same JA4T and JA4TS strings. Third, a tool's
database and the real system were different targets: over IPv4 Nmap reported
no exact match for the RHEL~9.9 host, and an exact match for a personality
made to follow its database. I describe the design and implementation of the system, the evaluation of
its deception capability, the measured overhead, and its limitations.
\end{abstract}


\section{Introduction}
Remote OS fingerprinting infers the operating system of a host from the way its
TCP/IP stack fills in fields that the specifications leave open. The idea of
defeating such probes is old~\cite{smart2000defeating,ippersonality,miyamoto2001aod}.
Misleading an adversary has since been listed as a resilience technique. NIST SP~800-160
Vol.~2~\cite{nist800160v2r1} lists among its cyber resiliency techniques
\emph{Deception}, defined as ``Mislead, confuse, hide critical assets from, or
expose covertly tainted assets to the adversary'', and \emph{Unpredictability},
defined as ``Make changes randomly or unpredictably'', with the aim to ``Keep the
adversary guessing.'' Against the reconnaissance technique Active Scanning it
names the candidate mitigation Present Deceptive Information, mapped to the
Disinformation approach and to control SC-30(4), Misleading Information
(its Table~F-3). A host that answers a scanner as a different operating system
can be read as a small, concrete instance of that mitigation.

Our 2001 paper~\cite{miyamoto2001aod} described an Anti OS Detection (\aod{})
system. Its implementation changed the FreeBSD~4.3 sources under \texttt{netinet} and was
evaluated with Nmap~2.54~BETA25, which predates the second-generation detection that
Nmap's documentation dates to 2006~\cite{nmaposdetect}. Three things have changed
since: the fingerprinting tools, the operating systems worth imitating, and the
kernel interfaces available to a module. This report asks what remains of the
original design under those changes.

\paragraph{Contributions.}
\begin{itemize}
\item An implementation as a loadable module for FreeBSD~15.1 that needs no
change to the kernel source and works over IPv4 and IPv6
(Section~\ref{sec:design}).
\item A comparison of the as-measured Linux personality and of the Windows
personality, each with the real guest it was built after, with an active tool
(Nmap's second-generation detection and its IPv6 detection) and with passive
ones (p0f~\cite{p0f3} and JA4T~\cite{ja4t} strings), over both protocols
(Sections~\ref{sec:nmap} and~\ref{sec:pair}).
\item The observation that Nmap's database and the RHEL~9.9 guest were
different targets (Section~\ref{sec:real}).
\item Initial sequence numbers and IP identification numbers of the
as-measured Linux and Windows personalities plotted against those of the
real hosts, in the form of the figures of 2001 (Section~\ref{sec:seq}).
\item Measurements of throughput, connection rate, ping time and the time
spent in the module's hooks over IPv4 and IPv6, the connection rate and the
hook times only while connections were opened and reset at once (Section~\ref{sec:perf}).
\item A comparison of the 2001 design with the present one, and of what
twenty-five years changed for both sides (Section~\ref{sec:years}).
\end{itemize}
The method itself is not new; Section~\ref{sec:related} places it.

\section{Related Work}\label{sec:related}
\paragraph{Removing the fingerprint on the path.}
The fingerprint scrubber~\cite{smart2000defeating,watson2004scrubbing}
normalizes traffic at a gateway. Mimura and Nakamura~\cite{mimura2008antiosfp}
modified the forwarding path of a Linux~2.6.18 router to reset the TTL,
randomize the IP ID, rewrite the TCP window (the stated value for the initial
window is 5840) and delete all TCP options from segments with the SYN flag
set, on packets forwarded from the inside to the outside; p0f~2.0.8 then reported the four hosts behind the router as
unknown, and Nmap~3.93 found no OS match for the one whose output is shown;
the authors state that it failed for the other three as well.
In their testbed inbound throughput fell from 15.98 to 5.79~MB/s for a Linux
host and from 12.08 to 5.87~MB/s for an OpenBSD host; the authors attribute
the first to the removal of the window scale option and the second to the
smaller initial window. They cite \aod{} and point out that it has to be
applied host by host and requires rewriting the kernel's TCP/IP stack, which
makes it difficult to apply to closed-source systems and appliances.

\paragraph{Answering as another system.}
IP Personality~\cite{ippersonality} did this on Linux as a netfilter target,
distributed as a patch to the kernel and to iptables; its change log runs
from a version of 2000 for a Linux~2.3.99 pre-release to the release of 2002
for Linux~2.4.18. It
tracked each connection so that it could imitate another system's sequence
number generator, limited the window, reordered options through a small
interpreted language, and answered Nmap's unusual probes itself. Its
documentation shows Nmap reporting the configured systems (Windows~9x/NT4,
AmigaOS) for the host itself, and no match for the one host behind it that it
shows; the output does not name the Nmap version, and the authors' own test
tool, shown in the section on development tools, reports an Nmap~2.53 backend. It also names
the risk I return to in Section~\ref{sec:years}: such tweaks ``could also make
the system weaker if the emulated IP stack is not as strong as the initial
one.'' Barroso~\cite{barroso2003practical} surveys the tools of that period. Honeyd~\cite{provos2004honeyd} answers for virtual hosts
and shapes their packets in a personality engine. IpMorph~\cite{prigent2010ipmorph} is a user-space TCP/IP stack with two sides,
placed between the scanner and the protected machine; it keeps per-flow state,
translates sequence numbers, and aggregates signatures of several tools
(Nmap, SinFP and Ring2 among them) into one personality; its p0f tests used configurations based on SinFP signatures. It already targeted Nmap's second-generation engine
(version 4.85~BETA~4) in 2009. The paper describes a test loop over the Nmap
database but reports no match rate and notes that Nmap did not recognize some
signatures, which the authors trace to a bug in Nmap; its p0f~2.0.8 and SinFP examples return a range of OpenBSD versions.
The paper reports no throughput or forwarding-latency benchmark for IpMorph. OSfooler-ng~\cite{osfoolerng} rewrites packets
in user space through a netfilter queue.
Yamada and Sato~\cite{yamada2017deception} answer Nmap's port, OS and
service-version scans with forged packets built with Scapy, a packet library
for Python, while a firewall discards the scan packets.
Albanese et al.~\cite{albanese2015deception} use Linux kernel modules on a netfilter hook for
outbound packets and target SinFP3; their own table lists the Nmap result as
unknown. SoFi~\cite{han2025sofi} generates adversarial examples under protocol
constraints so that a learned classifier assigns a packet to a wrong class. In its model it
treats DF, TTL, window size and flag bits as changeable, is evaluated against the authors'
own classifiers rather than against Nmap or p0f, and is by its own description
non-targeted: it aims ``to make OS fingerprints unidentifiable rather than
mimicking a specific OS.''

\paragraph{Position of this report.}
The closest prior work in mechanism is~\cite{albanese2015deception}; the
closest in ambition is IpMorph. Imitating a chosen system under Nmap's
second-generation detection is therefore not new. What this report adds is
narrower: the imitation is done inside the protected host by a
loadable module on FreeBSD, the results against Nmap~7.98, p0f~3 and JA4T
strings from my own script are set beside those of real hosts measured in
the same run, over IPv4 and IPv6, and throughput, connection rate, ping
time and hook times are measured over both. I
make no claim beyond that.

\section{Design}\label{sec:design}
\subsection{Why a packet hook is enough}
On FreeBSD's normal listener path the window and initial sequence number of a
SYN/ACK are produced by the SYN cache; replacing the
TCP function block would not reach them. I therefore register two
\texttt{pfil(9)} hooks on the IPv4 head, one inbound and one outbound, and the
same two on the IPv6 head.
On the outbound IPv4 side the IP header checksum is computed after the hook runs;
where the TCP checksum is still pending, a change that keeps the length of the
TCP header needs no checksum update (a shortened header gets a new
pseudo-header sum), and otherwise I apply an incremental update.

\subsection{What the hooks change}
\begin{itemize}
\item \textbf{Outbound:} TTL (where the packet carries the system default),
DF (on IPv4 TCP packets), IP ID sequence, window of SYN and SYN/ACK,
order of TCP options (under a personality that defines one, the list is rebuilt on eligible SYN and
SYN/ACK segments: the TCP header lies whole in the first buffer and carries a maximum segment size (MSS) option and, apart from NOP and end-of-list padding, none besides MSS, window scale,
SACK-permitted and timestamps; the list is written in place when its length is
unchanged, and a shorter one is installed only on a single-buffer segment without data whose
checksum is still pending), removal of the
timestamp option from the host's own SYN under the Windows personality where that rewrite applies, window scale value
(lowered only), the acknowledgment field of an RST with SYN, FIN and ACK clear, where the stack leaves it zero (Windows personality), ICMP echo reply code.
\item \textbf{Inbound:} under the Windows personality the ECE and CWR flags are cleared on a writable
SYN addressed to the host, with ACK and RST clear, to which none of the
answers named next applies; segments for which a personality defines another answer
(none of SYN, FIN, RST, ACK, PSH and URG set; SYN and FIN set with RST and
ACK clear; FIN, PSH or URG set with SYN, RST and ACK clear) are answered by the
module or silently dropped, as the personality prescribes.
\item \textbf{Both directions:} the sequence numbers of accepted connections
(Section~\ref{sec:isn}).
\item \textbf{One kernel variable:} per-connection timestamp offsets are turned
off for the database Linux and Windows personalities, so that FreeBSD uses
one offset for each pair of addresses, and the previous value is restored
afterwards.
\end{itemize}

Over IPv6 the hop limit takes the place of the TTL, the base header has no DF
bit and no IP ID, and the module leaves fragment headers alone. The two
outbound hooks share one TCP-shaping routine; separate inbound hooks apply
the TCP response rules for IPv4 and for IPv6. TCP is shaped only when
its header follows the IPv6 header directly; the hop-limit rule applies to eligible
locally generated packets regardless of the next header (forwarded packets,
packets marked for TCP segmentation offload (TSO) and packets whose IPv6 header the hook cannot write in place
keep their hop limit). The windows of SYN and SYN/ACK segments have their own
values over IPv6, where the real hosts sent other values than over IPv4.
For the Windows personality the module changes five more things over IPv6,
each after a reply of the Windows~11 guest:
\begin{itemize}
\item \textbf{Window.} On Nmap's probe without window scale, the Windows guest answered
with FED0 over IPv6 and FFDC over IPv4. These windows are consistent with
rounding 65535 down to a multiple of 1208 bytes over IPv6 and 524 bytes over
IPv4; the two observations do not establish the
rounding rule of Windows. The module's IPv6 value for a probe
with neither timestamps nor window scale (FC94) follows from that rule and was not measured on the Windows guest.
\item \textbf{Echo request with a nonzero code.} The Windows guest answered it with code
0. FreeBSD does not answer it over IPv6, so the inbound hook sets the
code of the request to 0 before the stack sees it. It does so for unicast requests
addressed to the host itself whose ICMPv6 header lies in the first, writable
buffer, directly after the IPv6
header or after one hop-by-hop header.
\item \textbf{Neighbor advertisement.} Answering Nmap's solicitation,
the Windows guest set the override flag and attached
its link-layer address. The
outbound hook adds both to a solicited advertisement that carries no option
and no extension header, on an interface with a six-byte link-layer address.
\item \textbf{RST to a bare ACK.} In the replies to Nmap's probes
\texttt{T4} and \texttt{T6} over IPv6, the acknowledgment field of the
Windows reply repeats its sequence field, and the module's reply does the
same.
\item \textbf{Parameter problem.} Answering Nmap's extension header probe,
the Windows guest sent a parameter problem message for an unrecognized next
header with pointer 40, where FreeBSD sends 56. The outbound hook sets the
pointer of such a message to 40.
\end{itemize}
For the Linux personalities the module changes two things over IPv6, each
after a reply of the RHEL~9.9 guest:
\begin{itemize}
\item \textbf{Echo request with a nonzero code.} The RHEL guest answered it
and repeated the code in its reply. The inbound hook builds that reply
itself and drops the request, under the same conditions as for the Windows
personality.
\item \textbf{Flow label.} The resets, ICMPv6 errors and echo replies of the
RHEL guest carried a flow label; FreeBSD sends resets and ICMPv6 errors with
label~0. The outbound hook puts a label, a keyed hash of the addresses, on
TCP segments, ICMPv6 errors and echo replies that have none.
\end{itemize}
Under the other personalities the flow label is left as FreeBSD sends it,
and the replies that the module builds itself carry flow label~0.

\subsection{Initial sequence numbers}\label{sec:isn}
On FreeBSD the initial sequence number of a SYN/ACK comes from the SYN cache,
which replacing the TCP function block does not reach. Under the Linux and
Windows personalities the outbound hook therefore computes a number in the
way of the imitated system, remembers the difference from the stack's own
number for that connection, and applies it to the later segments of the
connection: it adds the difference to the sequence numbers the host sends
and subtracts it from the acknowledgment numbers and SACK blocks that come
back. The stack keeps its own numbers throughout. The differences are kept
in a table with one entry per connection, which is the only per-connection
state of the module. Two restrictions apply: a segment is translated only
when its TCP header lies whole and writable in the first buffer, directly
after the IP header, and a SACK option only when it starts at an even offset
in the TCP header.

\paragraph{When a connection is over.}
A hook sees packets, not the state of the stack, so the module cannot know
for certain when a connection has ended. It therefore never removes an entry
because it saw a reset or a FIN. It only marks the connection as over. The
mark has two uses, both described below: an entry that is over can be given up
before others when the table is full, and a mark set by a reset decides how a later
SYN/ACK between the same addresses and ports is read.

A reset that the host sends sets the mark. A reset that arrives sets it only
when its sequence number equals the acknowledgment number the host sent
last; with its default settings FreeBSD makes the same comparison before it
drops a connection that is established or being
accepted~\cite{rfc5961}. The mark of a reset is taken back when the stack
then sends a segment on the connection that is not a reset. A connection is also over when a
FIN has passed in each direction and each has been acknowledged.

When the host sends a SYN/ACK between addresses and ports that already have
an entry, the module has to tell a retransmission from a new connection. It
compares two numbers of the segment, as the stack wrote them, with two
numbers kept in the entry: the sequence number with that of the SYN/ACK that
made the entry, and the acknowledgment number with the one the host sent
last. If both pairs are equal and the entry is not marked by a reset, the
module takes the segment for a retransmission and translates it with the
difference already in the entry. Otherwise it takes the segment for the
start of a new connection, which reuses the entry with a newly computed
number; if the personality selected at that moment does not imitate
sequence numbers, the entry is removed and the connection keeps FreeBSD's
number.

The replies that the module builds itself (the inbound hook, above) are not
segments of the stack and must not be translated. The module attaches a tag
to each and does not translate the sequence number of a tagged segment;
when no memory is free for the tag, the reply is sent without one.

\begin{figure}[t]\centering
\includegraphics[width=0.62\linewidth]{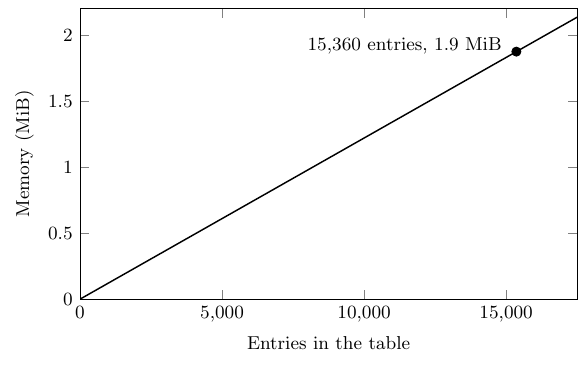}
\caption{Memory of a full table of sequence numbers by the number of its
entries, calculated as 128 bytes per entry; nothing in this figure was
measured. The mark is the size of the module, 15,360 entries.}\label{fig:tablemem}
\end{figure}

\paragraph{The table.}
The module keeps one entry of about 128 bytes per accepted connection. The
size of such a table is a trade-off against memory: a larger table remembers
more connections and takes more memory (Figure~\ref{fig:tablemem}). I set
it to 15,360 entries (about 1.9~MiB), the default size of FreeBSD's TCP SYN
cache. The SYN cache holds connections only while they are being accepted,
so this is a choice of a number that FreeBSD already uses for a table of
connections, not one derived from a requirement. When a
new connection needs an entry and the table is full, the module gives up one
entry, the first that it finds in this order:
\begin{enumerate}
\item an entry marked by a reset, at once; or an entry whose connection
closed with FINs and on which no segment has passed for more than
60~seconds (twice the maximum segment lifetime that FreeBSD assumes by
default);
\item an entry on which the module has seen no segment arrive with an
acknowledgment and without a reset, and on which no segment has passed for more than 30~seconds;
\item any other entry on which no segment has passed for more than
300~seconds.
\end{enumerate}
The entries are kept on
three queues, one for each case, so the search looks at the head of each
queue and not at the whole table. If none of the three heads qualifies, no
entry is given up and the new connection keeps the stack's own number.

\paragraph{The imitated number.}
The imitated number is a keyed hash of the two addresses and the two ports
plus a clock. For Linux the clock counts once every 64~ns, as in the Linux
source (\texttt{net/core/secure\_seq.c}). For Windows~11 I have no source. The module counts once every
3~microseconds, that is, 333,333 per second, a value I chose to be close to
the growth of my guest (Section~\ref{sec:seq}).
The hash is a simple integer mixing function with a random key chosen when
the module is loaded, not a cryptographic one. An entry is made when the
host sends a SYN/ACK. A connection that the host opens in the ordinary way, in
which it sends a SYN and no SYN/ACK, keeps FreeBSD's numbers; if an entry is
left between the same addresses and ports, it is removed when that SYN
passes.

\subsection{Personalities}
A personality is a table entry, selected by one setting
(\texttt{net.inet.tcp.emu\_kind}). Three are evaluated here. The
\emph{database Linux} personality follows Nmap's database; the
\emph{as-measured Linux} personality follows a Red Hat Enterprise Linux~9.9
host; the \emph{Windows} personality follows a Windows~11 host. The two Linux
personalities differ in one thing: the database one turns off FreeBSD's
per-connection timestamp offsets, which changes the result of Nmap's
timestamp test, and the as-measured one leaves them on.
Following the paper of 2001, I write d(OS) for an \aod{} system with the
personality of that OS selected: d(Linux) is the system with the as-measured
Linux personality, and d(Windows) the system with the Windows personality.
The personalities of 2001 are carried over only in part (Section~\ref{sec:years}).
Tables~\ref{tab:linux} and~\ref{tab:win} list the differences between
FreeBSD~15.1 and each target that the module addresses over IPv4, and which
part of the module closes each gap.

\begin{table}[t]\centering\small
\caption{Linux personalities. WS is the window scale option and TS the
timestamp option. For the row on the SYN+FIN+PSH+URG segment, the
FreeBSD column gives what I observed with Nmap's probe to the open port. The
last column says which part of the module
makes the change: the outbound hook rewrites packets that the host sends, the
inbound hook handles packets that arrive, and ``kernel variable'' is the
setting of the timestamp offsets (Section~\ref{sec:design}).}\label{tab:linux}
\begin{tabular}{llll}\toprule
Field & FreeBSD 15.1 & Linux target & Changed by\\\midrule
Option order & MSS, NOP, WS, SACK, TS & MSS, SACK, TS, NOP, WS & outbound hook\\
Window scale & 8 & 7 & outbound hook\\
SYN/ACK window & FFFF & FE88 with TS, FAF0 without & outbound hook\\
Nmap's timestamp test & \texttt{TS=20} to \texttt{22} & \texttt{TS=A} (database personality) & kernel variable\\
SYN+FIN+PSH+URG & answered with SYN/ACK & no answer & inbound hook\\
IP ID, TCP except SYN/ACK, RST & 0 when DF is set & nonzero & outbound hook\\
IP ID, ICMP echo reply & FreeBSD's & grows by a random step & outbound hook\\
Initial sequence number & FreeBSD's & hash plus a 64~ns clock & both hooks\\\bottomrule
\end{tabular}
\end{table}

\begin{table}[t]\centering\small
\caption{Windows personality. For the rows on the segment with no flags and
on the SYN+FIN+PSH+URG segment, the FreeBSD column gives what I observed with
Nmap's probes to the open port. The last column is as in
Table~\ref{tab:linux}. $^{*}$A value I chose (Section~\ref{sec:isn}).}\label{tab:win}
\begin{tabular}{llll}\toprule
Field & FreeBSD 15.1 & Windows target & Changed by\\\midrule
TTL & 64 & 128 & outbound hook\\
IP ID & 0 when DF is set & incremental (Nmap: \texttt{TI=I}) & outbound hook\\
SYN/ACK window & FFFF & FFDC on Nmap's probe without WS & outbound hook\\
ECN in the reply to Nmap's probe & \texttt{CC=Y} & \texttt{CC=N} & inbound hook\\
Padding without TS & SACK, EOL, EOL & NOP, NOP, SACK & outbound hook\\
Timestamps on the host's own SYN & offered & not offered & outbound hook\\
Nmap's timestamp test & \texttt{TS=20} to \texttt{22} & \texttt{TS=A} & kernel variable\\
Segment with no flags & no answer & RST+ACK & inbound hook\\
SYN+FIN+PSH+URG & answered with SYN/ACK & RST+ACK & inbound hook\\
RST to a bare ACK & ack field 0 & ack field nonzero & outbound hook\\
ICMP echo reply code & echoed & 0 & outbound hook\\
Initial sequence number & FreeBSD's & hash plus a 3~\textmu s clock$^{*}$ & both hooks\\\bottomrule
\end{tabular}
\end{table}

\section{Evaluation}
All results in this section come from one run
of the measurement scripts, made by the author on 1 December 2026 with the
module as published. The raw output, the scripts and the checksums of the
source and module files are in the repository
(\texttt{test/results/official-20261201b}).

The personalities were adjusted on the two reference guests described below
until the output of the tools agreed. The comparisons in this section
therefore show that the adjustment was completed for these guests and tools;
they are not a test on a host or with a tool that the module was not built
against. The numbers of repetitions (one Nmap scan of each target, 200 scans
in Section~\ref{sec:tries}, 100 probes in each series of
Section~\ref{sec:seq}, five rounds, 5000 and 40,000 connections, 3000 pings)
are my choices, fixed in the scripts before this run. None was derived from
a required precision.

Target: FreeBSD 15.1-RELEASE-p7 in a QEMU/KVM guest (8 vCPU, 8~GB), with port
22 open and 23 closed. Prober: the physical machine that runs the guests
(Linux), over a dedicated virtual link
that is not forwarded anywhere. Two reference systems ran as guests on the
same machine, each on a link of its own: Red Hat Enterprise Linux~9.9 (RHEL~9.9 below; kernel
5.14.0-741.20.1.el9\_9, 2~GB, with TCP timestamps and window scaling on) and Windows~11 (an evaluation image; when the guest was set up on 21 November
2026 I recorded the Enterprise edition, build 22621, and turned its
firewall off). On the RHEL guest ports 22 and 23 were used as on the FreeBSD
guest; on the Windows guest port 135 was open and 23 closed.
Each target was scanned once with \texttt{nmap -O -d -n -Pn --osscan-guess}
and the two ports over IPv4, and once with the same options and \texttt{-6}
over a link-local address. One
run of Nmap is not one measurement: in the source of Nmap~7.98, the IPv4
detection makes up to five tries by default, makes no further try once one
has succeeded with a perfect match, and bases its verdict on the best try;
the fingerprint it prints holds the tries together. Where it finds no perfect
match, Nmap run with these options can print guesses with a percentage. Over
IPv4 that percentage is a share of points, rounded down: every attribute of
a test that has a value both in the measured fingerprint and in the database
entry carries points, and they are won when the measured value satisfies the
entry. It is not a probability. Over IPv6 a separate classifier
gives every label a score between 0 and 1, which Nmap prints as a percentage
with a guess; it prints a label as its result when one score stands clear of
the others and the measured features are close enough to that label's
training data.
For the passive tools, the prober opened three connections to an open port of
each target over each protocol, and each target made three attempts to
connect to a closed port of the prober over each protocol (the Windows guest
made one over IPv6); all were captured. p0f~3.09b read the captures, and my own script
computed JA4T strings for the SYN segments and JA4TS strings for the SYN/ACK
segments. The script was not compared with the reference implementation and
does not measure the retransmission timing that the reference JA4TS can
append.

\subsection{Verdicts}\label{sec:nmap}
\begin{table}[t]\centering\footnotesize
\caption{Verdicts for FreeBSD without a personality, for the module's
personalities and for the real systems.
``Exact'' means that Nmap printed the label as its result and did not mark it
as a guess; it does not mean that the replies are identical. Nmap identifies
IPv6 hosts with a separate classifier, and its labels differ from those over
IPv4; for Windows they name older systems. The JA4T value for Windows equals
the one the JA4 repository lists for Windows~11~\cite{ja4t}. $^{*}$A single
Nmap scan can report several exact matches; in this scan OpenWrt~21.02 and
MikroTik RouterOS 7.2--7.5, which are Linux-based, follow Linux 4.15--5.19 (Section~\ref{sec:tries}).}\label{tab:results}
\begin{tabular}{p{2.3cm}p{3.5cm}p{3.1cm}p{1.6cm}l}\toprule
Host & Nmap 7.98, IPv4 & Nmap 7.98, IPv6 & p0f (SYN, IPv4) & JA4T (SYN, IPv4)\\\midrule
FreeBSD, personality off & no exact match (first guess Android 4.0.1--4.0.4, 93\%; then FreeBSD 12.0--13.0, 92\%) & no match (guess: FreeBSD 12.X, 91\%) & FreeBSD (fuzzy) & \texttt{65535\_2-1-3-4-8\_1460\_8}\\\midrule
Linux personality (database) & exact: Linux 4.15--5.19$^{*}$ & exact: Linux 4.19 & Linux 2.2.x--3.x & same as RHEL~9.9\\
Linux personality (as measured) & no exact match; first guess Linux 5.4--5.10 (97\%) & exact: Linux 4.19 & Linux 2.2.x--3.x & same as RHEL~9.9\\
RHEL 9.9 & no exact match; first guess Linux 5.4--5.10 (97\%) & exact: Linux 4.19 & Linux 2.2.x--3.x & \texttt{64240\_2-4-8-1-3\_1460\_7}\\\midrule
Windows personality & exact: Windows 10 1703 or Windows 11 21H2 & exact: Windows Vista, 7 SP1, Server 2008 R2 or 8.1 & Windows NT kernel & same as Windows~11\\
Windows 11 & exact: Windows 10 1703 or Windows 11 21H2 & exact: Windows Vista, 7 SP1, Server 2008 R2 or 8.1 & Windows NT kernel & \texttt{64240\_2-1-3-1-1-4\_1460\_8}\\\bottomrule
\end{tabular}
\end{table}

As my script computes it, a JA4T string joins four values of a SYN segment
with underscores: the
window size, the kinds of its TCP options in the order in which they appear,
the MSS, and the window scale. The options are written
as their kind numbers: 1 is NOP (padding), 2 MSS, 3 window scale, 4
SACK-permitted and 8 timestamps. The systems differ in the window, in
the order of the options and in the window scale, and the Windows string has
no 8 because my Windows~11 guest did not offer timestamps on the SYN
segments it sent. My script builds a JA4TS string in the same way from a SYN/ACK segment.

With the personality off, Nmap~7.98 does not identify FreeBSD~15.1:
its first guess over IPv4 is Android. p0f labels the SYN as FreeBSD, as a
fuzzy match. The as-measured Linux personality has the row
of the RHEL host, and the Windows personality the row of the Windows host.

\subsection{Each personality against the real host}\label{sec:pair}
A shared verdict can hide different replies, so I compared the fingerprints
themselves (Table~\ref{tab:pair}).

\begin{table}[t]\centering\small
\caption{Each personality beside the real host it follows.}\label{tab:pair}
\begin{tabular}{p{6.2cm}p{4.0cm}p{4.0cm}}\toprule
 & d(Linux) and RHEL~9.9 & d(Windows) and Windows~11\\\midrule
Nmap verdict, IPv4 & same & same\\
Tries inside the Nmap run, IPv4 & 5 and 5 & 1 and 1\\
The twelve test lines from \texttt{OPS} to \texttt{IE}, IPv4 & all the same & all the same\\
Sequence classes \texttt{TI}, \texttt{CI}, \texttt{II}, \texttt{SS} & \texttt{TI} and \texttt{CI} the same and \texttt{SS} absent from both; \texttt{II} missing in one of the five tries on the personality and in two on the RHEL host & same\\
Nmap's timestamp test \texttt{TS} & 20 to 22 over the five tries on the personality; 21 in all five on the RHEL host & A and A\\
Nmap verdict, IPv6 & same & same\\
p0f signature of SYN and of SYN/ACK, IPv4 and IPv6 & same & same\\
JA4T and JA4TS strings, IPv4 and IPv6 & same & same\\
TTL and hop limit & 64 and 64 & 128 and 128\\\bottomrule
\end{tabular}
\end{table}

The twelve test lines are those of Nmap's IPv4 fingerprint whose values do
not depend on sampled numbers (\texttt{OPS}, \texttt{WIN}, \texttt{ECN},
\texttt{T1} to \texttt{T7}, \texttt{U1}, \texttt{IE}). The remaining line,
\texttt{SEQ}, holds the sequence classes and the timestamp test of
Table~\ref{tab:pair} and three values that Nmap computes from six sampled
sequence numbers (\texttt{SP}, \texttt{ISR}, \texttt{GCD}); these three
differ from try to try on a real host as well. Nmap printed the \texttt{SEQ} line of every try.
Table~\ref{tab:seq} gives the three values.

\begin{table}[t]\centering
\caption{The three values of the \texttt{SEQ} line that Nmap computes from
six sampled sequence numbers, in hexadecimal as Nmap prints them. They
differ from try to try on one host, so the rows are to be compared as
ranges, not value by value.}\label{tab:seq}
\begin{tabular}{lrrr}\toprule
 & \texttt{SP} & \texttt{ISR} & \texttt{GCD}\\\midrule
Linux personality (as measured), five tries & FB--107 & 108--10E & 1\\
RHEL 9.9, five tries & FC--10A & 107--10C & 1\\\midrule
Windows personality, one try & 108 & 109 & 1\\
Windows 11, one try & FF & 10E & 1\\\bottomrule
\end{tabular}
\end{table}

Over IPv6, p0f gave every host the label ``???'', so for IPv6 I compared
the signature strings that p0f printed and not its labels. p0f does handle IPv6: in
the source of p0f~3.09b, a signature that names no IP version is also matched
against IPv6 segments, with its IPv4 flags (DF, IP ID) set aside. But a
segment with a nonzero flow label gets the flag \texttt{flow}, a segment
with that flag does not match a signature that lacks it, and no signature in
the database carries it. Every IPv6 SYN and SYN/ACK in my captures, from
the real hosts and from the module alike, had a nonzero flow label.

Over IPv6, Nmap's fingerprint records the replies themselves. Each
personality sent the same replies as its host, with the same TCP flags,
windows, option layouts, hop limits, ICMPv6 types, codes and pointers, and
with a flow label where the host sent one. Sequence numbers, the values of
the flow labels and checksums differ.

\subsection{The database and the real system}\label{sec:real}
The RHEL~9.9 guest did not match the database. Nmap made five tries on
the RHEL~9.9 guest, found no perfect match and printed guesses, the first of
them Linux 5.4--5.10 at 97\%. The as-measured personality got the same
result. The database personality differs from it in one setting, the
timestamp offsets: with one offset for each pair of addresses, Nmap's
timestamp test gives \texttt{TS=A}, a value the database entry accepts,
instead of the \texttt{TS=21} that the tries on the RHEL host gave. With that
change alone Nmap reported an exact match on its first try, under the label
Linux 4.15--5.19.

Matching a tool's database and matching the real system are different goals,
so the module keeps both Linux personalities; I did not study which of
them serves a defender better. For the Windows~11 guest the two goals coincided: it was
an exact match.

p0f~3.09b did not name the current systems precisely either: it reported the
RHEL guest as ``Linux 2.2.x--3.x'' and gave the SYN/ACK segments of the real
hosts no label. For p0f I therefore took as the goal the verdict that the
real system gets.

\subsection{Initial sequence numbers and IP identification}\label{sec:seq}
\begin{figure}[t]\centering
\begin{minipage}[t]{0.49\linewidth}\centering
\includegraphics[width=\linewidth]{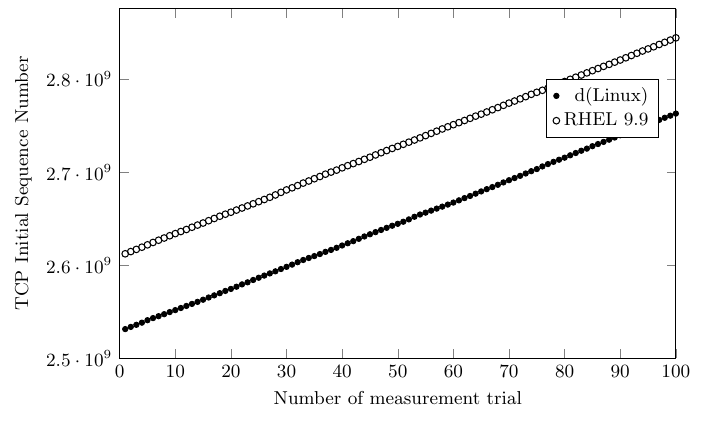}
\caption{Initial sequence numbers: the as-measured Linux personality,
d(Linux), and the RHEL~9.9 host.}\label{fig:isslinux}
\end{minipage}\hfill
\begin{minipage}[t]{0.49\linewidth}\centering
\includegraphics[width=\linewidth]{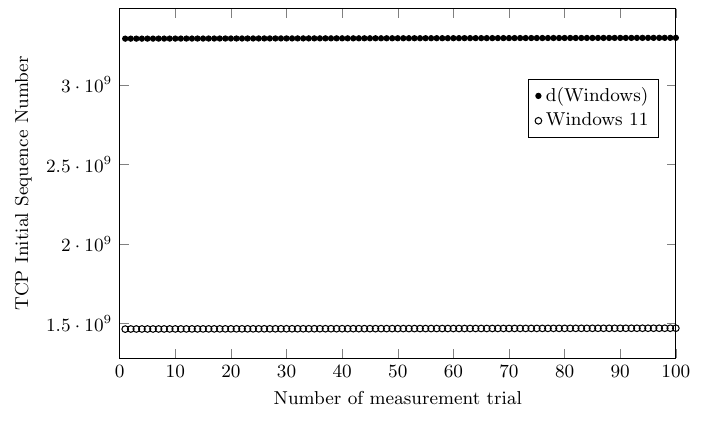}
\caption{Initial sequence numbers: the Windows personality, d(Windows), and
the Windows~11 host. At this scale the growth of both (5 to $6 \times 10^6$
over the series) cannot be seen; Table~\ref{tab:iss} gives the growth per
second. The first step of the Windows host and the uneven growth described
in the text cannot be seen here either.}\label{fig:isswin}
\end{minipage}
\end{figure}

\begin{figure}[t]\centering
\begin{minipage}[t]{0.49\linewidth}\centering
\includegraphics[width=\linewidth]{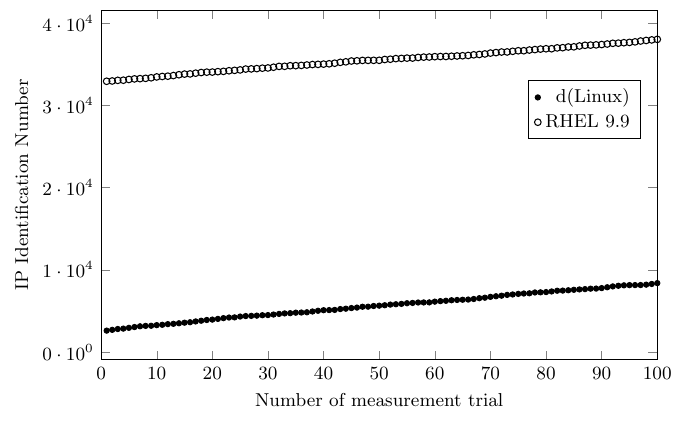}
\caption{IP ID of echo replies: the as-measured Linux personality and the
RHEL~9.9 host.}\label{fig:idlinux}
\end{minipage}\hfill
\begin{minipage}[t]{0.49\linewidth}\centering
\includegraphics[width=\linewidth]{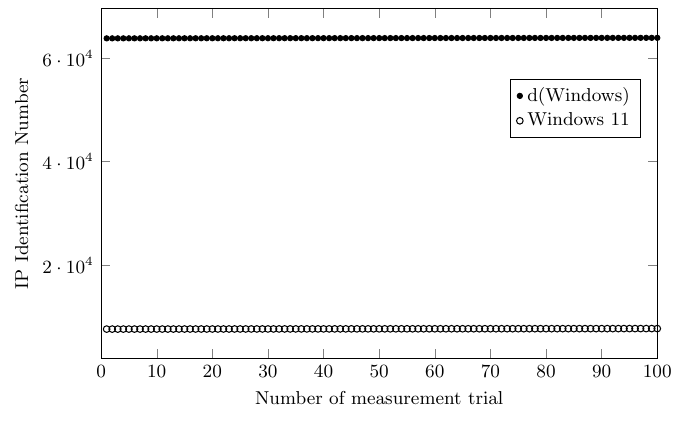}
\caption{IP ID of echo replies: the Windows personality and the Windows~11
host. Both grow by 1 per reply, which this scale does not show.}\label{fig:idwin}
\end{minipage}
\end{figure}

The 2001 paper plotted the initial sequence numbers of its personalities
against those of the real systems. Figures~\ref{fig:isslinux}
and~\ref{fig:isswin} do the same for the present module, and
Figures~\ref{fig:idlinux} and~\ref{fig:idwin} do it for the IP ID. For the sequence
numbers, the prober sent 100 SYN segments to an open port from one source
port, about 0.15~s apart, recorded the sequence number of each SYN/ACK and
reset the half-open connection. In a second series it sent 100 SYN segments,
each from another source port, about 0.1~s apart.
For the IP ID, it sent 100 echo requests about 0.1~s apart and recorded the
IP ID of each reply.


\begin{table}[t]\centering
\caption{Initial sequence numbers on one source port, 100 probes. The growth
per second is the increase from the first number to the last, modulo
$2^{32}$ at each step, divided by the time between them. The last column
gives the smallest and the largest of the 99 values obtained by dividing
each step by the time it took. $^{*}$The first step of the Windows host was
644,566 (4,127,600 per second); its other 98 steps were 43,298 to 59,364,
which is 338,177 per second together and 276,061 to 397,826 per second step
by step.}\label{tab:iss}
\begin{tabular}{lrr}\toprule
 & growth per second & growth per second of single steps\\\midrule
Linux personality (as measured) & 15,623,617 & 15,620,406--15,626,758\\
RHEL 9.9 & 15,623,576 & 15,619,164--15,627,933\\\midrule
Windows personality & 333,304 & 333,242--333,375\\
Windows 11 & 378,236$^{*}$ & 276,061--4,127,600$^{*}$\\\bottomrule
\end{tabular}
\end{table}

The starting values differ. The clock of the Linux personality is the 64~ns
clock of the Linux source, and its rate and that of the RHEL host are within
0.001\% of each other (Table~\ref{tab:iss}). The clock of the Windows
personality counts 333,333 per second, a value I chose
(Section~\ref{sec:isn}). The first step of the Windows host was 644,566,
more than ten times its largest other step. I do not know its cause. The personality
makes no such step. Over its other 98 steps the
Windows host grew by 338,177
per second, 1.5\% faster than the personality; with the first step it grew
by 378,236 per second, 13.5\% faster.

The two differ in more than the first step. A step divided by the time it
took is 333,242 to 333,375 per second for the personality, whose numbers on
one source port change only with its clock, and 276,061 to 397,826 per
second for the Windows host over its 98 later steps. An observer who records
the times can tell the two apart.

From changing source ports the numbers of all four went up in about
half of the steps (37 to 53 of 99, counting a step as upward when the
difference modulo $2^{32}$ is below $2^{31}$) and down in the others.
The IP ID of the echo replies grew by 3 to 104 per reply (mean 58.4) under
the Linux personality and by 2 to 104 (mean 51.5) on the RHEL host; it grew
by exactly 1 under the Windows personality and on the Windows host.
The script that sent the echo requests also opened and closed 100
connections to the open port and made 100 attempts on the closed port, about
0.1~s apart, and recorded the IP ID of each SYN/ACK and of each reset. On
the SYN/ACK segments it was 0 on every one from the Linux personality and
from the RHEL host; it grew by 3 per segment on the Windows host and by 4 to
6 under the Windows personality. On the resets it was 0 from the Linux
personality and from the RHEL host and grew by 1 per reset under the Windows
personality and on the Windows host. The open port was 135 on the Windows
host and 22 on the FreeBSD guest, and this record does not tell me whether
the difference on the SYN/ACK segments comes from the ports or from the
module.

\subsection{How often Nmap settles on its first try}\label{sec:tries}
Because Nmap measures again when a try is not a perfect match, I ran the
same Nmap command 200 times for each of the two personalities that get an
exact match. The number 200 is my choice. For the database Linux
personality all 200 scans reported Linux 4.15--5.19 as an exact match on the
first try; the further entries varied: OpenWrt~21.02 and MikroTik
RouterOS~7.2--7.5 followed in 152 scans, OpenWrt~21.02 alone in 26, MikroTik
RouterOS alone in 3, Linux~6.0 in one, and no further entry in 18. In 68 of
these 200 fingerprints the sequence class \texttt{II} was missing. For the
Windows personality all 200 scans reported the same exact match, 195 on the
first try and 5 on the second.
These are the counts of one series. The real hosts and the as-measured Linux
personality were not scanned 200 times, so I have nothing to compare these
counts with.

\subsection{Overhead}\label{sec:perf}
\begin{table}[t]\centering\small
\caption{Medians of five rounds; in each round the five conditions were
measured in the order of the rows. In each round: iperf3 for 10~s in each
direction; 5000 connections opened one after another and reset at once,
without data, the rate being 5000 divided by the total time, with the median
and the 99th percentile of the time to open one connection; and 3000 pings,
of which the mean round-trip time is taken. Receiving and sending are seen
from the guest.}\label{tab:perf}
\begin{tabular}{llrrrrrr}\toprule
 & Condition & recv Gbit/s & send Gbit/s & conn/s & connect \textmu s & 99\% \textmu s & ping ms\\\midrule
IPv4 & no module & 2.56 & 17.54 & 12079 & 55.8 & 110.2 & 0.106\\
 & loaded, off & 2.45 & 17.47 & 13240 & 55.7 & 106.1 & 0.108\\
 & Linux (database) & 2.49 & 17.89 & 14590 & 50.0 & 105.9 & 0.105\\
 & Linux (as measured) & 2.69 & 18.60 & 13958 & 52.7 & 127.2 & 0.106\\
 & Windows & 2.75 & 17.77 & 13814 & 54.1 & 105.4 & 0.105\\\midrule
IPv6 & no module & 2.51 & 17.91 & 11397 & 63.7 & 114.9 & 0.111\\
 & loaded, off & 2.56 & 17.87 & 14949 & 48.6 & 89.2 & 0.119\\
 & Linux (database) & 2.50 & 18.71 & 13058 & 53.6 & 97.2 & 0.116\\
 & Linux (as measured) & 2.71 & 17.51 & 12345 & 58.9 & 149.1 & 0.120\\
 & Windows & 2.56 & 17.11 & 13803 & 50.6 & 147.4 & 0.115\\\bottomrule
\end{tabular}
\end{table}
Table~\ref{tab:perf} gives the medians of six quantities from the records of the script. All of these measurements were made between the physical machine and a
virtual machine running on it, and the machine is shared with other work; I have no bare-metal numbers.
The ranges of the five rounds with the module overlap those without it in
every column (for the connection rate over IPv4 without the module, 10,858
to 14,687 per second), with one exception: over IPv6, under the database
Linux personality and under the Windows personality, all five rounds had a
higher connection rate and a shorter median connect time than any round
without the module (12,568 to 15,821 and 12,804 to 14,806 connections per
second against 10,507 to 11,837). The medians with a personality selected
lie on both sides of those without the module. From these numbers I can
neither show nor exclude an overhead of the module. Within each round the conditions were measured
in the same order, with the condition without the module first, and I did
not vary that order.

To see the module's own work I timed its four hook functions with DTrace
while the prober opened and at once reset 40,000 connections, without data,
under each condition
(Table~\ref{tab:hook}). The times include that of the DTrace probes, which
I did not measure separately. With the personality off a hook shapes
nothing and only translates the connections that still have an entry. With a personality selected, the mean time of a call was 71 to 259~ns more
than with the personality off over IPv4, and 258 to 425~ns more over IPv6.
Each condition was timed once, in a fixed order, so I cannot give the
variation of a single timing. The two Linux personalities run the same hook
code and differ only in a kernel variable. Their times differ by 188~ns
inbound over IPv4 (1107 and 1295) and by 65 to 79~ns in the other three
columns. While the hooks were being timed, the prober opened 9,163
connections per second over IPv4 with the personality off and 9,204 to
10,452 with a personality, and 10,521 over IPv6 with the personality off and
7,876 to 11,292 with a personality.

\begin{table}[t]\centering
\caption{Mean time per hook call in nanoseconds (DTrace; about 120,000
inbound and 40,000 outbound calls per condition and protocol).}\label{tab:hook}
\begin{tabular}{lrrrr}\toprule
 & IPv4 in & IPv4 out & IPv6 in & IPv6 out\\\midrule
loaded, off & 1036 & 887 & 857 & 832\\
Linux (database) & 1107 & 1057 & 1115 & 1192\\
Linux (as measured) & 1295 & 1130 & 1194 & 1257\\
Windows & 1243 & 1098 & 1134 & 1234\\\bottomrule
\end{tabular}
\end{table}

The connection rate of Table~\ref{tab:perf} and the hook times of
Table~\ref{tab:hook} were measured while the prober opened connections and
reset them at once. That is a condition I chose, and it matters for the sequence
numbers: the entry of a connection that ended with a reset can be given up
at once, while the entry of a connection that closed with FINs is not given up to
make room for more than 60~seconds (Section~\ref{sec:isn}). On a connection that closes with FINs the module also records the state of
the closing: the number that acknowledges each FIN, and whether each
acknowledgment has passed. The connections that the prober opened for the connection rate and for the
hook times carried no FIN and no data, so these two measurements tell
nothing about that work, about the translation of data segments, or about a
table that holds many entries of connections closed with FINs. Those cases are not among the measurements reported here.

\section{Twenty-five years}\label{sec:years}
Table~\ref{tab:years} sets the 2001 implementation beside the present module.

\begin{table}[t]\centering\small
\caption{The 2001 implementation and the 2026 module.}\label{tab:years}
\begin{tabular}{p{2.9cm}p{5.6cm}p{5.6cm}}\toprule
 & 2001 (paper; later source tree) & 2026\\\midrule
Form & changes to the \texttt{netinet} sources of FreeBSD~4.3 and one build option in \texttt{conf/options}; ten changed \texttt{netinet} files in the surviving tree & loadable module, two \texttt{pfil} hooks for each of IPv4 and IPv6, kernel source untouched\\
Protocols & evaluated over IPv4 & IPv4 and IPv6\\
Tools considered & Nmap~2.54~BETA25 in the paper; Xprobe and the Netcraft survey are named in the later source & Nmap~7.98 (second generation, and its IPv6 detection), p0f~3, JA4T\\
Systems imitated & NetBSD~1.4, Windows~95/98/NT~4.0 in the paper; Windows~2000 and Cisco IOS also in the later source tree & Linux and Windows~11 evaluated; the 2001 personalities are kept in part\\
Initial sequence numbers & generator rewritten per system & the stack's numbers translated by a difference kept per connection in a table of 15,360 entries, for Linux and Windows~11\\
SYN/ACK window & a fixed value for the NetBSD and Windows personalities in the FreeBSD~4.3 patch; in the later source tree some values also depend on the MSS and window scale options of the peer's SYN & chosen from a small table by the timestamp and window scale options that the hook parses in the outgoing SYN/ACK\\
Unusual flag combinations & handled inside \texttt{tcp\_input} & answered or dropped in the inbound hook\\
RST rate limit & in the later source tree, bypassed in the TCP input path when built with \texttt{TCP\_EMUOS} & left alone; the replies that the module sends itself do not pass through it\\
ICMP & quoted UDP checksum and, under the Windows personality, the IP ID in the FreeBSD~4.3 patch; in the later source tree also echo reply, address mask reply, TOS, TTL, DF & echo reply code, TTL, IP ID; over IPv6 the hop limit, the code of an inbound echo request, and the neighbor advertisement\\
Reference & Nmap's verdict, \texttt{tcpdump} output, and observation of real systems for sequence numbers & the tools' databases and real hosts measured in the same run\\\bottomrule
\end{tabular}
\end{table}

\paragraph{Sequence numbers, from outside the kernel.}
The 2001 paper treated the initial sequence number as a fingerprint,
and the 2001 \aod{} system replaced the generator for its
personalities. That is a local change inside the kernel source. From outside it, a
packet hook that changes the first sequence number must translate every later
sequence number of the connection and the acknowledgments coming back; SACK blocks and the TCP header quoted in
ICMP errors refer to the same numbers. IP Personality~\cite{ippersonality}
translated sequence and acknowledgment numbers with an offset kept per
connection, IpMorph~\cite{prigent2010ipmorph} translated sequence numbers in
user space, and FreeBSD's own \texttt{pf} also rewrites SACK blocks for its
sequence modulation. The present module translates sequence numbers, acknowledgment numbers and
SACK blocks for the Linux and Windows personalities
(Section~\ref{sec:isn}); it does not translate the header quoted in an ICMP
error. The 2001 change needed no table of connections for this. The module keeps a table of
fixed size and has to judge from the packets when a connection is over
(Section~\ref{sec:isn}).

What is imitated has changed as well. The generator that the 2001 paper
describes for its Windows personality was predictable, and reproducing such
a generator would weaken the host instead of only changing its appearance,
as the IP Personality authors already warned. The Linux source adds a clock
to a keyed hash of the addresses and ports, so that without the key the
numbers seen on one pair of addresses and ports say nothing about those on
another. The module follows the Linux form for both personalities, but its hash is a
simple mixing function (Section~\ref{sec:isn}), and I regard it as a
demonstration, not as a generator to rely on.
The 2001 personalities are another matter. I kept them only in part. The FreeBSD one
shapes nothing. For the other four the module sets
their TTL, DF bit, IP ID step, fixed SYN and SYN/ACK windows, their reply to a
segment without flags and, where the later source tree has them, the SYN+FIN drop
and echo reply code 0. It leaves out the sequence number generators, the
option changes, the further window rules (such as those that depend on the
peer's SYN), the replies to FIN, PSH and URG segments and the other UDP and
ICMP changes of that tree. They were not evaluated for this report.

\paragraph{What changed around the module.}
Mimura and Nakamura~\cite{mimura2008antiosfp} objected that \aod{}
needs a rewritten kernel. With \texttt{pfil} hooks and checksum offloading, the
header fields of Section~\ref{sec:design} can now be changed by a loadable module. The guidance changed too: the 2001 approach can be read as an instance of
what is now listed as a resilience technique~\cite{nist800160v2r1}. The tools changed too.
Over IPv4, Nmap~7.98 gave no exact match for the RHEL~9.9
guest I tested; p0f~3.09b
labelled it Linux 2.2.x--3.x; JA4T reduces the SYN to one line of four
fields. For the RHEL~9.9 guest the database and the real
system disagreed.

\paragraph{What did not change.}
In the 2008 testbed, inbound transfers to the two hosts
measured behind the router lost between half and two thirds of their
throughput, which the authors attribute to the removed window scale
option and the smaller initial window. I did not measure this.
And the approach still reaches only as far
as the fields I thought of.

\section{Limitations}
\begin{itemize}
\item The TTL or hop limit is set on each locally sent packet that the module
handles and that carries the system default (forwarded packets are not
touched), and the IP ID on each IPv4 packet for which the
personality defines one. The DF setting of the 2001 personalities (on IPv4 TCP packets), the
acknowledgment field of a bare RST and the ICMP changes are likewise not
limited to handshakes. Windows and options are shaped
only on SYN and SYN/ACK segments. IPv4 packets that reach the hook as
fragments, packets marked for TSO at the outbound hook over either protocol,
and packets whose IP header that hook cannot write in place are not shaped.
Packets marked for TSO thus retain the stack's IPv4 TTL and
IP ID or its IPv6 hop limit; the translation of sequence numbers does not
depend on that mark. I did not capture a large transfer to measure how
often this occurred. The module does not touch FreeBSD's retransmission
timers or congestion control, or anything above TCP.
\item Initial sequence numbers are imitated only for connections on which
the host sends a SYN/ACK, and only under the Linux and Windows
personalities; a connection that the host opens in the ordinary way gets
FreeBSD's numbers. The hash is not a cryptographic one. The Windows
personality was compared with one guest only. After its first step, the growth per second of that guest
varied from step to step between 276,061 and 397,826, that of the
personality between 333,242 and 333,375, so an observer who records the times can tell
the two apart (Section~\ref{sec:seq}). The first
step of that guest in the series was more than ten times its others; the
personality makes no such step.
\item The TCP header quoted in an ICMP error is not translated, and a SACK
option is translated only when it starts at an even offset in the TCP
header. IPv4 segments that reach a hook as fragments are not translated. A
translated connection stops working when the module is unloaded.
Connections that got an imitated number keep it after the personality is
switched off.
\item A translated connection also stops working when its entry in the table
is given up. Apart from a new connection between the same addresses and
ports, an entry is given up only when the table of 15,360 entries is full
and a new connection needs an entry (Section~\ref{sec:isn}); a connection on
which no segment has passed for more than 300~seconds can lose its entry in
this way while it is still open. What the module knows of a connection it
reads from the packets that pass, and this can differ from the state of the
stack; such a difference changes which entry is given up first. A connection
that arrives while no entry can be given up keeps FreeBSD's numbers. The size of the table is fixed when the module is
built.
\item The connection rate and the hook times were measured while the prober
opened connections and reset them at once, a condition I chose; the overhead on
connections that close with FINs, whose entries are not given up to make
room for more than 60~seconds, is not among the measurements reported here (Section~\ref{sec:perf}). Whether 15,360
entries are enough depends on the connections a host receives; this report
contains no measurement of that.
\item The Linux personalities draw the IP IDs of the IPv4 TCP segments that
the module shapes, other than SYN/ACK and RST segments, from one counter per
network stack.
\item Over IPv6, TCP is shaped only when its header follows the IPv6 header
directly. The flow label is FreeBSD's where FreeBSD sets one; under the Linux
personalities the module puts one on resets, ICMPv6 errors and echo replies
that would have none.
\item The window scale can be lowered but not raised.
\item My criterion is the output of Nmap, p0f and my JA4T script, checked
against one RHEL~9.9 guest and one Windows~11 guest; the Windows guest had
its firewall turned off, which is not its default setting. A tool that looks at
other fields may still tell the difference, and a real host of another
version or configuration may answer differently from the ones I measured.
\item The module can be detected. As the IpMorph authors say of their own tool,
``it is quite easy to detect the presence'' of such software; a host whose
kernel is identified as Windows while its applications are plainly not is one
example.
\item A personality is static. Changing it at random times would add the
Unpredictability technique of~\cite{nist800160v2r1}; switching is a single
\texttt{sysctl} write, but I have not studied its effect on open connections.
\item I have not tested against the tabular Transformer classifiers
of~\cite{perezjove2025tabular} or against Hershel~\cite{shamsi2016hershel}.
\end{itemize}

\section{Conclusion}
For the targets and tools I evaluated, the goal of the 2001 design was reached by a
loadable module of about 1,700 lines of C and an 82-line header, with no
kernel patch. Measured in one run beside the two guests on which the personalities had
been adjusted, the as-measured Linux personality and the RHEL~9.9 guest, and
the Windows personality and the Windows~11 guest, received the same verdicts
from Nmap over IPv4 and IPv6, the same p0f signatures and the same JA4T and
JA4TS strings over both protocols; this shows that the adjustment was
completed for these guests and tools. The initial sequence numbers of the
Windows guest began with a step more than ten times its others, and after
it their growth per second varied from step to step between 276,061 and 397,826 where that of the personality stayed between 333,242 and 333,375; an
observer who records the times can tell the two apart. Measured between the physical machine and a virtual machine on it, the ranges of
throughput, connection rate and ping time with the module overlapped those
without it or lay on the faster side, so I could neither show nor exclude
an overhead of the module. The connection rate and the hook times were measured while the prober opened
connections and reset them at once. The mean time of a hook call, including
the DTrace probes that timed it, was then 0.8 to 1.3 microseconds; with a
personality selected it was 71 to 259~ns more over IPv4 and 258 to 425~ns
more over IPv6. Each condition was timed once, and two conditions that run
the same hook code differed by 65 to 188~ns. Connections that
close with FINs are not among the measurements reported here. The imitated sequence numbers need a table of
connections, which the 2001 change did not: 15,360 entries of about 128
bytes each, about 1.9~MiB; this report contains no measurement of whether
that size is enough for a given host. For the RHEL~9.9 guest, Nmap's database and
the real system turned out to be different targets: over IPv4 Nmap found no exact match
for that guest and an exact match for a personality that differs in one
setting from the one built after it.

\paragraph{Availability.} The module, the measurement scripts and the raw
output of the measurements are released under the BSD 2-Clause license at
\url{https://github.com/daisu-mi/aod}.

\section*{Use of AI tools}
The author used Claude Code (Anthropic), running the model Claude
Mythos~5.2, to rebuild \aod{} as a FreeBSD loadable module, drawing on the
historical patch for the 2001 personalities and on Nmap fingerprints and host
measurements for the current ones; to write the measurement scripts; and to prepare the text
of this manuscript together with the author. The
measurements reported here were run by the author with those scripts. The
kernel patch and the paper of 2001 predate these tools. The author set the
design and its requirements, chose the reference systems, decided how the
module was to be evaluated, ran the measurements reported here, and reviewed
and corrected the implementation and the text. The author is responsible for
the whole content.

\bibliographystyle{unsrt}
\bibliography{refs}
\end{document}